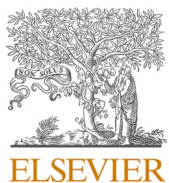
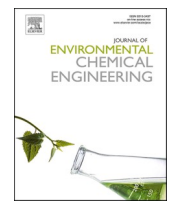

# Plasma water treatment for PFAS: Study of degradation of perfluorinated substances and their byproducts by using cold atmospheric pressure plasma jet

Barbara Topolovec [a,b], Olivera Jovanovic [c], Nevena Puac [c], Nikola Skoro [c], Elisabeth Cuervo Lumbaque [a], Mira Petrovic [a,d,*]

[a] Catalan Institute for Water Research (ICRA), Emili Grahit 101, Girona 17003, Spain
[b] University of Girona, Girona, Spain
[c] Institute of Physics, University of Belgrade, Pregrevica 118, Belgrade 11080, Serbia
[d] Catalan Institution for Research and Advanced Studies (ICREA), Passeig Lluis Companys 23, Barcelona 08010, Spain

ABSTRACT

This study evaluates the effectiveness of non-thermal plasma at atmospheric pressure (NTP APPJ) for treating PFAS - contaminated water in different matrices. Successful removal of several perfluoroalkyl carboxylic acids (PFCAs) (C6 to C4), perfluroalkane sulfonic acids (PFSAs) (C8 to C4) and perfluropolyethers (PFPEs) (GenX and ADONA) PFAS compounds was achieved in laboratory scale experiments. In the deionized water (DW), high removal efficiencies (> 90%) were observed for longer-chain PFAS, PFOS (99.89%), PFHxA (94.61%) and ADONA (94.83%), while shorter- chain compounds had lower removal rates. Real water samples (tap water and synthetic effluent) showed lower overall degradation percentages (8–50%) depending on compound and matrix. Short-chain PFAS displayed around 10% removal in tap water, while PFOS and GenX achieved 50% and 32% removal, respectively. Complex matrix effects influence degradation rates. Byproducts from the plasma treatment were investigated, revealing distinct degradation mechanisms for various PFAS compounds. For PFSAs and PFCAs, degradation involved electron transfer, bond breaking and subsequent reactions. Conversely, ADONA and GenX degradation initiated with ether-group cleavage, followed by additional transformation processes. Plasma-based technology shows potential for degradation of PFAS, especially for newer substitute compounds like ADONA and GenX. However, further research is needed to optimize plasma performance for complete mineralization of PFAS. This study also proposes a degradation mechanism for ADONA, marking a novel investigation into ether-group PFAS degradation with potential implications for further research and understanding toxicological implications.

## 1. Introduction

The dispersive occurrence of per – and polyfluorinated substances (PFAS) in the environment and the impact on human and ecological health are raising concern and social awareness in the last few decades [2,52]. These organic molecules, so called "forever chemicals" are part of a man-made production since the 1950 s and are used on everyday basis for commercial and consumer purposes. PFAS, as a complex group of substances with over 4700 identified chemicals, have varying properties, but one thing in common – incredibly strong and stable carbon-fluorine bond which is very difficult to break down. Fluorinated carbon chain or "tail" can vary in length chain; therefore, substances are sometimes divided into short-chain and long-chain PFAS. On the one end (or both) polar or ionizable "head" group is attached which also can be variable, like carboxylic and sulfonic acid [8]. This structure provides them with unique physicochemical properties, high thermal and chemical stability, repellent to water, oils, and grease. Because of such properties and variability in structure, they can be used for diverse application, such as electronics manufacturing, aqueous film-forming foams (AFFF), dust suppressions, hydraulic fluids, textile stain repellents, food containers, cleaning products, polishes, waxes, paints, cosmetics, etc. Barzen-Hanson et al., [3,37,49].

The reports on PFAS presence in drinking water near chemical plants and blood serum samples of humans and animals from the areas that are close to fluoropolymer production facilities are known for a few decades, but not until studies about findings of PFAS in remote areas, certain global concern increased and PFAS have been started to be monitored worldwide [4,34,38]. Those were one of the first proves that these substances are very persistent and even though the most important input of PFAS to the environment comes from direct release from industrial facilities, and indirect source is also affective, from different precursors as degradation products, with long-range transport potential. Other studies had shown that their bioaccumulation in drinking water sources and potential to accumulate in the dietary products and humans are related to serious health effects, such as immune suppression, thyroid disease, cancer, significant (chronic) toxicity of some [20,25,33]. Since 2009, one of the most known compounds to be used and present in the environment, perfluorooctane sulfonate (PFOS) and perfluorooctanoic acid (PFOA) were recognized as potentially dangerous compounds for human health and the environment, and have been regulated throughout United States and Europe [12,45]. In 2022, The European Commission adopted the proposal to revise the list of priority substances in surface water and groundwater where it stays that sum of 24 PFAS the groundwater quality standards are 0.004 μg/L [13]. Therefore, global production of PFOA and PFOS started to be restricted or banned while new fluorinated substances, as replacement, were developed [47]. These new substances are mostly part of short-chain PFAS group, such as perfluorobutane sulfonic acid (PFBS) with C4 chain-length. Another interesting group are perfluropolyethers (PFPEs), or shortly ether-PFAS, which include (among others) substances named hexafluoropropylene oxide dimer acid (HFPO-DA;GenX), dodecafluoro-3 H-4,8-dioxanonanoate (ADONA) and 6:2 chlorinated polyfluoroalkyl ether sulfonate (6:2 Cl-PFESA; F-53B). More detailed discussion can be found in [47] and [30]. While, because of the alkyl-ether (C-O-C) addition in their chemical structure, higher mobility properties and less bioaccumulation can be expected, environmental impact and toxicity is still not certain, yet concerning. Recent studies shows that even though they do not have a high potency to accumulate in aquatic species, their presence in water (drinking and groundwater) has more impact on animals and people that are consuming it and are exposed to them [51]. [31] has reported GenX presence in human serum and fish muscle while ADONA has been detected in blood donor samples in Germany [14]. Levels of ADONA in surface water and its toxicity have been investigated by [17], while GenX has been found in rivers of The Netherlands and Germany [15,21]. Global distribution of these substances has also been reviewed by [32]. Toxicological data can be found in [7] as well.

In terms of wastewater treatment processes, many studies and practice had shown that most PFAS are highly recalcitrant to conventional wastewater treatment, but also to many advanced oxidation processes (AOP). Adsorption on activated carbon and ion exchange methods are the most common water treatment technologies that have been used for PFAS removal from water media [9,10,50]. However, the search for more efficient technologies that do not require high energy input, waste management and long treatment time is needed. Few AOPs have proven to be effective in their removal from water, such as electrochemical oxidation [11], sonolysis [16], photocatalysis ([26]) and among them, plasma treatment technologies with somewhat different configurations of plasma sources than one presented here in this paper [35,36,41]. Non-equilibrium or non-thermal plasma (NTP) is based on the production of highly reactive oxygen and nitrogen species which are generally needed for an effective water treatment, namely atomic hydrogen and oxygen, hydroxyl and hydroperoxyl radical, nitrogen-species, aqueous electrons etc. [1,5,6]. The main advantage of this technology is that only the energy applied to the working gas (air, agon, helium, mixtures etc.) is needed for plasma generation. Since plasma contains such a rich and reactive chemistry, it became of great interest in terms of use for removal of very recalcitrant compounds [22, 28,48]. In our recent review paper, we have summarized the current knowledge on the chemistry and degradation pathways of OMPs by using different NTP types, including PFAS [44]. [42] was investigating degradation of PFOA and PFOS substances in groundwater samples. [39] was using landfill leachate in experiments for the investigation of several different PFAS substances with plasma jet above liquid with formation of bubbles. While current studies (and more that have been mentioned in review paper) shown high removal efficiency and promising results in reactors of around 1 L, it should be noted that most of the studies have been investigating PFOS and PFOA substances and up to this date according to the authors' knowledge, a similar investigation has not been done in terms of plasma treatment of fewer known compounds, ADONA and GenX. A few of them mentioned more complex matrices but with still limited information on PFAS byproducts after plasma treatment. [40] had done interesting work in breakdown products of PFAS in plasma-based water treatment processes, PFOA and PFOS compounds.

In this work, NTP in gas-liquid environment generated with pin-type electrode setup at a laboratory scale has been used for the removal of six substances which were selected as a representative of groups of perfluorinated compounds, carboxylates, sulfonates, and ethers, in order to study plasma application on different groups of PFAS. The removal efficiency of each substance, individually, was assessed in 3 different matrices: distilled water (DW), tap water (TW), wastewater or secondary effluent (SE), for the same experimental conditions. The influence of NTP treatment on different working parameters was also observed. In addition to that, in all cases, liquid phase byproducts were studied using HRMS, with more attention on novel substances such as ADONA. The first attempt of proposed structures of studied byproducts of GenX and ADONA as well as degradation mechanism of ADONA have been given.

## 2. Materials and methods

### 2.1. Chemicals and standards

Standards for substances that were used in experiments, perfluorooctanesulfonic (PFOS), nonfluorobutane-1-sulfonic acid (PFBS), perfluorobutyric acid (PFBA), undecafluorohexanoic acid (PFHxA) were purchased from Sigma Aldrich. Hexafluoropropylene acid (HFPO - DA or GenX) and dodecafluoro-3 H-4,8,-dioxanoate (ADONA) were purchased from Wellington laboratories. All solutions were prepared in Mili-Q water at room temperature ($\pm$ 25°C). High purity methanol and Mili-Q water used for mass spectrometry analysis were purchased from Thermo Fischer Scientific, while ammonium acetate used for mobile phases was obtained from VWR. The chemicals used were of either analytical grade or high purity (>98%). Details of the compounds such as structures, molar mass etc. can be found in Table S1 of Supporting Information (SI).

### 2.2. Analytical methods

The analytical method used for target PFAS analysis was based on the method reported from [27] with a few alterations. A 5500 QTRAP hybrid triple quadrupole-linear ion trap mass spectrometer (QqLIT-MS) with a turbo Ion Spray source (Applied Biosystems, Foster City, CA, USA) coupled to a Waters Acquity Ultra-Performance™ liquid chromatograph (UPLC) (Milford, MA, USA) was used. For the chromatographic separation, an Acquity UPLC C18 column (50 mm×2.1 mm, 1.7 μm) was used coupled with Phenomenex Luna 5 μm precolumn C8 (50 mm×3 mm, 100 Å) to minimize contamination from mobile phases. The chromatographic separation used aqueous solutions of ammonium acetate (5 mM) (A) and methanol (B). The established flow rate was 0.4 mL min$^{-1}$, and the elution gradient condition was finalized at 90% of B and 10% of A. The sample injection volume was 10 μL, and the limit of detection (LOD) for this method was determined to be 0.1 μg L$^{-1}$. Table S2 (SI) summarizes the optimized compound-dependent parameters of the MS.

Liquid chromatography – High resolution Mass Spectrometry (LC-HRMS) methodology using Orbitrap Exploris 120 (OE120) was performed to identify possible degradation byproducts of all compounds. EPA Method 57 was adapted from the EPA Thermo Scientific source with some minor modifications. As a mobile phases, 5 mM ammonium acetate (A) in water and methanol (B) were used. Final conditions consisted of 90% (A) and 10% (B) with a flow rate of 0,4 mL min$^{-1}$ and pressure of 195 bar. The injection volume was 5 μL. The data was processed using the Compound Discoverer (CD) software (Text S2(SI)) from Thermo Scientific, which is designed for non-target screening of all types of organic compounds. General instrument parameters can be found in Table S3 (SI).

### 2.3. Plasma treatment of PFAS setup

#### 2.3.1. Plasma source configuration

The treatments of PFAS-polluted water samples were performed using a non-thermal plasma (NTP) treatment with an atmospheric pressure plasma jet (APPJ) source. Electrical characterization of this device is presented in [23] and [24]. Laboratory-scale APPJ source configuration is shown schematically in Scheme 1 Figure S1. The APPJ consisted of a stainless-steel high-voltage (HV) electrode (1 mm in diameter) inserted inside a ceramic tube, which was then inserted in a glass tube with outer and inner diameters of 6 and 4 mm, respectively. The electrode was connected to a commercial HV RF power source (T&C Power Conversion AG0201HV) with a frequency of 332 kHz. In the experiments, argon was used as a working gas, with a flow rate of 1 standard liter per minute (slm). The argon flow rate was controlled by mass flow meter (OMEGA, FMA5800/5500). Time-variable current and voltage signals were monitored with a Tektronix oscilloscope (MDO3024 model), and a HV probe (Tektronix 6015 A) was used to determine the voltage at the HV electrode. The glass sample vessel ($\varphi=$ 55°mm) had copper tape at the bottom and it was grounded via a resistor of $R=$ 1 kΩ for current monitoring. The streamer type discharge was generated above the liquid in a 10 mm gap between the tip of the powered electrode and the surface of the liquid sample.

#### 2.3.2. Water samples treatment experiments

Individual solutions of PFOS, PFBS, PFBA, PFHxA, GenX, ADONA were prepared in different types of water matrices: distilled water (DW), tap water (TW) and secondary effluent (SE) with initial concentration of 100 μgL$^{-1}$. In all cases, 10 mL of solution was subjected to the plasma treatment for different time intervals of 3, 5, 7 and 10 min. After the plasma treatment, samples were taken to investigate the degradation profile through target analysis and for the study of byproducts through non-target analysis. Measurements of pH, conductivity, and temperature for both control samples and each plasma – treated sample were conducted. Control samples consisted of untreated solutions in DW, TW and SE. DW served as a model to explore optimal removal conditions without matrix interference. TW and SW allowed us to assess PFAS behavior and degradation efficiency in more complex environments. Additionally, the impact of plasma treatment on various parameters under the same initial concentration and treatment duration as in DW were examined. Furthermore, the volume of all samples was measured after each treatment.

## 3. Results and discussion

### 3.1. Degradation of PFAS in distilled water

The removal efficiency of each investigated PFAS in DW during 10 min of plasma treatment is shown in Fig. 1.

As it can be seen, after 10 min of treatment time, removal efficiency of PFOS, PFHxA and ADONA reached over 90%. The removal efficiency trend for all compounds is as follows: PFOS (C8) > PFHxA (C6) > ADONA (C7) > PFBA (C4) > GenX (C6) > PFBS (C4). It can be observed that the longer the chain-length of the compound, the higher the efficiency for the same treatment time. As a result of the structure of PFAS compounds, surface activity increases with chain length, which means that long-chain PFAS will have better contact with the plasma at the plasma-liquid interface and faster degradation. Short-chain PFAS tend to accumulate less at the surface therefore longer treatment time is needed. The fastest degradation occurred at the beginning (between 3 and 6 min) for all studied PFAS, followed by a slow increase. It is possible that slow degradation is due to the presence of byproducts and depending on the compound, those byproducts are produced in different treatment times. The trend of rapid degradation is more pronounced for three compounds with longer chain length (PFOS, ADONA, PFHxA) when compared to GenX, PFBA and PFBS. The lowest degradation was observed for PFBS with only around 30% removal efficiency in 10 min. For the same treatment time, it took around 57% for PFBA, which has the same chain length as PFBS. Compared to sulfonates, carboxylates are harder to degrade which has been shown in previous studies [39,41]. Most of the results in this paper showed the same consistency, only not in the case of PFBS and PFBA. However, it seems that the hydrophobicity of PFBA is slightly higher than PFBS when we take their values of *logKow* into account which can explain higher removal efficiency of PFBA. According to the results, high removal efficiency occurred in the case of ADONA (around 95% in 10 min) while for GenX was 50% for the same treatment time. From the compound properties (Table S1, SI), it can be seen that ADONA has $CF_3O$ extra group which can give higher surface activity to the molecule.

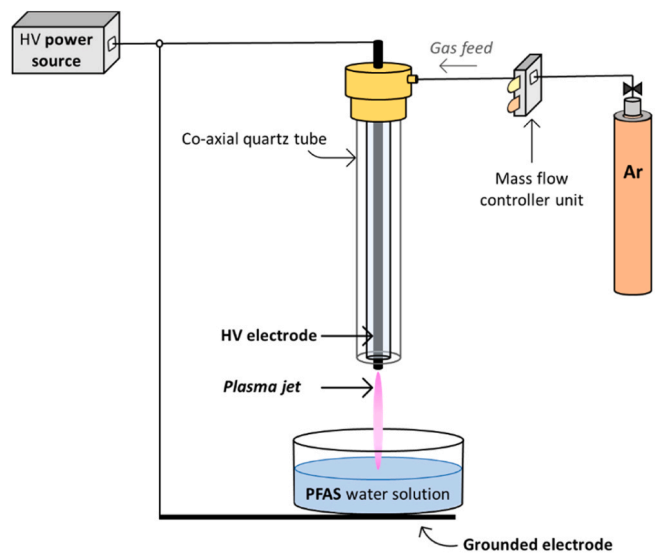

**Scheme 1.** Configuration of plasma system.

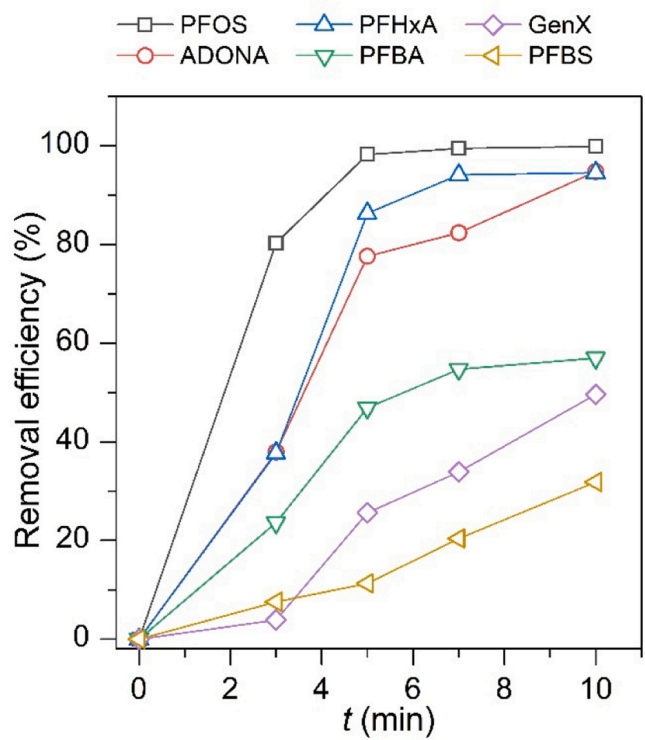

**Fig. 1.** Removal efficiency of PFAS in DW.

A few parameters were measured during the treatment of the solutions. Fig. 2a and b show the pH, conductivity, and temperature measurement results (average values). Additionally, the treated water volume was also measured at the end of each treatment. The same change in pH was observed for each compound in all experiments. The initial pH was between 6 and 7.5, while after 10 min of treatment time, the pH value decreased to around 3. The pH decreased drastically at the beginning of the treatment (in the first two min), and then varied around 3 for the rest of the treatment. The reduction in the pH value could be attributed to a generation of some reactive species. Although argon was used as a working gas, some nitrogen species could occur in the solution due to plasma contact with ambient air [24,29]. Additionally, byproducts resulting from the degradation such as sulfuric acids and the production of few other plasma species, $H^+$ and hydrogen peroxide, could contribute to an acidic environment and lead to decrease in pH [29].

The measured initial conductivity of solutions was around 4 μS cm$^{-1}$ and it increased linearly in all experiments. After 10 min of treatment time, it was around 400 μS cm$^{-1}$. The trend suggests that with longer treatment time, conductivity is expected to increase. What was indicated in some previous work ([39,41,46]) in plasma-liquid systems, conductivity of water tends to increase due to interaction with electrons formed in the discharge which can break water molecules into ions and ionic species produced in plasma are transferred into liquid, hence, more ions are present in water. While some authors indicated negative correlation between the degradation rate of PFAS and extremely high conductivity (above 20 mS/cm) assuming high concentration of ions reduced the contact area with contaminated water [41], other indicated that conductivity did not affect the degradation in significant way or even have a positive impact [46]. Therefore, for precise assessment it should be considered which type of discharge was involved and what the level of conductivity was. In our experiments, the degradation rate for all compounds decreased over time and this was probably caused by the competition between byproducts and parent compound to react with plasma species and not due to the changes in conductivity which reached 400 μS/cm and could be considered as low.

Volume of the samples was measured after the treatment to observe possible loss due to evaporation. The volume loss after 10 min treatment time was 10% and this was taken into account by normalizing the data for all samples. To investigate the evaporation we measured temperature in the bulk of the solutions. The temperature of the solution gradually increased under plasma exposure, with a maximum of 45°C in some cases. The increase in temperature over treatment time indicates that surface processes were dominantly governing the evaporation and not the heating of the sample volume. One of the possible mechanisms was the local heating at the point of contact of the plasma streamer and water due to the higher local temperature (still below 100°C). Another mechanism was the change in the partial pressure in the interface gas/liquid region due to argon flow and the plasma. This localized change in the partial pressure leads to evaporation of the treated liquid sample. The rise in the temperature, as shown in Fig. 2, is not linear and the heating slows down never going above 50°C, which indicates that degradation of contaminants is purely due to non-thermal processes.

### 3.2. Degradation of PFAS in complex matrices

In terms of PFAS degradation in NTP treatment – gas-liquid interface, studies so far have shown promising results, but most of them were conducted in pure water matrices, with PFOS and PFOA as representatives. However, in some recent study [18] PFAS removal from groundwater (already contaminated with PFAS) using plasma has been investigated. Singh et al., [39] has investigated PFAS removal by using landfill leachate sampled and showed the potential of plasma technology for treatment of PFAS contaminated water. In this work, the removal efficiency in TW and SE has been assessed for all 6 substances as in DW, as a function of the treatment time. Results have been given in Fig. 3a and b.

In both type of matrices, lower removal efficiency was observed in comparison to experiments in pure water. Short-chain PFAS (PFBS, PFBA and PFHxA) showed almost linear degradation and very low percentage of removal efficiency (less than 10%) with respect to PFOS, GenX and ADONA. A major difference was in the case of PFHxA where efficiency was around 94% in DW while in TW and SE it was around 6% and 9%, respectively. With the complexity of the matrix, the removal efficiency is changing, especially for compounds that do not easily migrate to the liquid surface.

Fig. 4a. shows trend of the pH (average values) of the TW and SE samples during NTP treatment. The initial pH for both TW and SE matrices were around 8.3 and this value did not significantly change with plasma treatment even for the longest treatment time of 10 min. It should be noted that pH is not necessarily dependent on the main

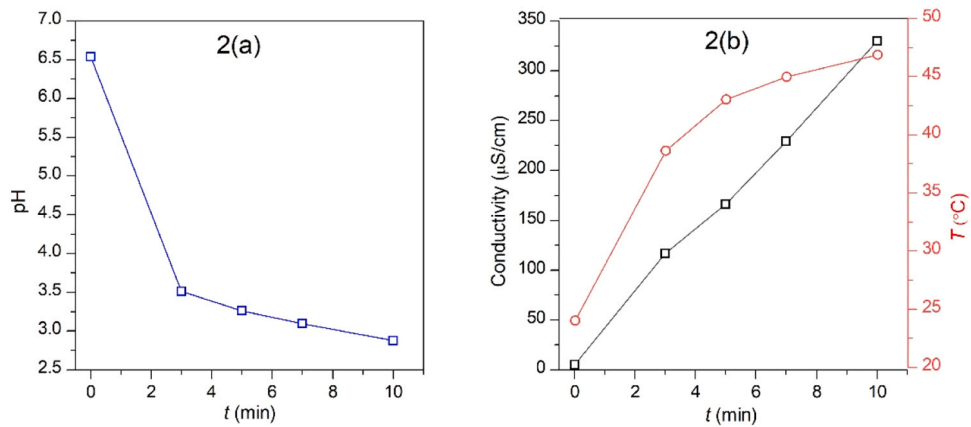

**Fig. 2. (a) and (b).** Dependency of pH, conductivity and temperature during treatment time of PFAS (total average value).

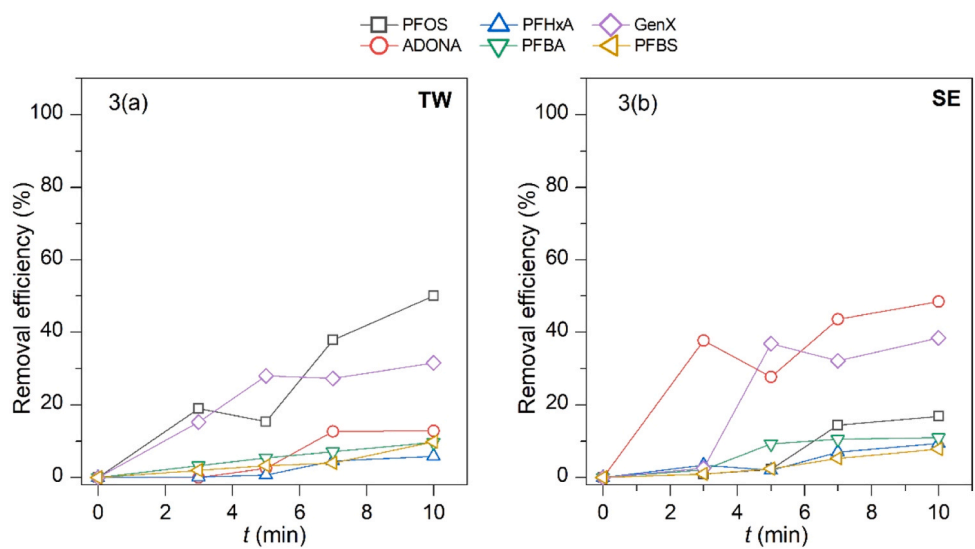

**Fig. 3. (a) and (b).** Removal efficiency of PFAS in TW and SE.

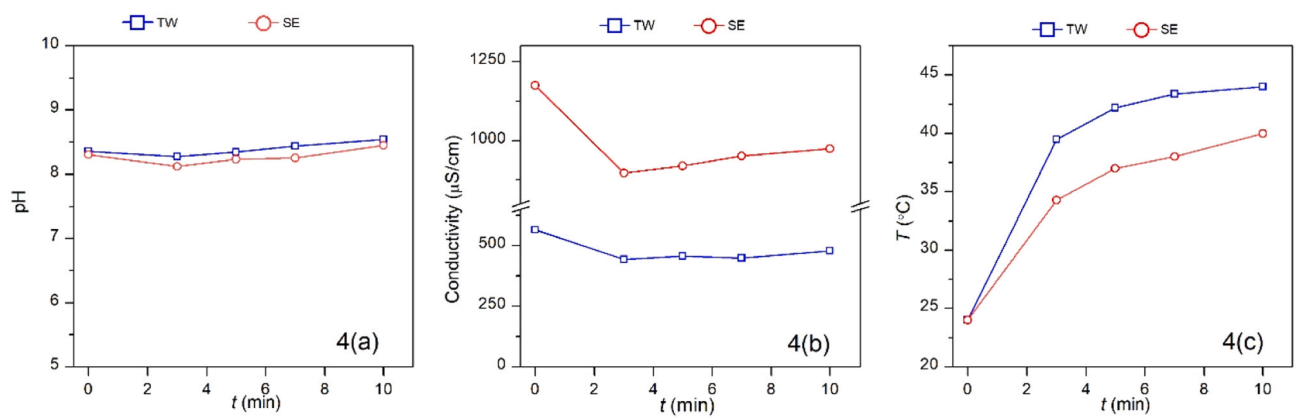

**Fig. 4. (a), (b) and (c).** Dependency of pH, conductivity and temperature during treatment time of PFAS in TW and SE (averaged values obtained from all samples).

compounds that were investigated. In fact, the pH did not drastically change when PFAS compounds were added to all solutions. Considering the chemistry of water samples, it is possible that in real water samples the concentration of basic products such as carbonate anions calcium and potassium (which where not present in DW) is much higher compared to the expected acid products of PFAS, hence the neutral to base value of pH.

The changes of the solutions conductivity (Fig. 4b) are also diverse when compared to DW. Generally, the conductivity was higher in TW (initial around 500 μS/cm) and SE (initial around 1200 μS/cm) than in DW (5 μS/cm). In the case of all compounds, the trend shows a small decrease at the beginning of the treatment followed with a slight increase in the end of the treatment, in both TW and SE. However, in the case of SE, the decrease is more significant than in TW. Since the solutions are much more complex from the chemical point of view, the matrix definitely impacts their conductivity. Also, it can be concluded that plasma treatment did not significantly affect the water conductivity in real water samples. As for the temperature, it can be seen (Fig. 4c) that the trend was very similar to the one in DW. At the end of the treatment, the maximum volume loss was around 7%, which indicates that the same conclusion related to evaporation mechanisms could be drawn as in the case of DW samples.

### 3.3. PFAS byproducts

Few previous studies discussed possible byproducts of PFOA and PFOS by utilizing different plasma configurations, as well as proposed degradation pathways [19,40,42,43]. In this work, the analysis of the transformation products (TP) was performed by using the LC-Orbitrap HRMS for all compounds in all 3 matrices. In total, 6 different byproducts have been found. Table 1. shows proposed structures and observed mass of byproducts as well as their parent compounds and type of matrices where they are detected. TP_212 has been also reported in [40], while others are reported for the first time in this paper.

#### 3.3.1. Carboxylates and sulfonates byproducts

The cold atmospheric pressure plasma jet (CAP) generates highly reactive species, including $e^-_{aq}$, $^\bullet OH$, and $^\bullet H$. In water with a neutral pH, PFOS exists as its anion species, capable of swiftly reacting with $e^-_{aq}$ to create the corresponding radical anions. Subsequently, the reactive species could induce the cleavage of C-C, resulting in the formation of TP_448 (*m/z* 449.9333). Also, the subsequent reactions with $e^-_{aq}$ and/or $^\bullet OH$ can led to the formation of perfluoric acid intermediates, such as TP_212 (*m/z* 212.9793) or PFBA. Both have been detected after 10 min of treatment time. According to the previous literature, degradation of long-chain PFAS during plasma treatment involves interaction of argon ions, aqueous and plasma electrons and hydroxyl radicals which lead to the formation of short-chain PFAS. PFBA is known as a common byproduct of PFOS, and based on this hypothesis, it should degrade further by losing another (-CF$_2$) group. Then, the interaction with reactive species degrades into short chain organic acids with the possibility to reach mineralization. In this case, it can be concluded that indeed electrons and argon ions play an important role in initiation reaction by attacking the C-S bond, followed by propagation reaction that result in the formation of short – chain PFBA. Longer treatment time was needed to detect whether PFBA is present after 10 min and/or it was degraded to some level. In DW samples, no byproducts were detected, which means either PFOS was completely degraded or the level of byproducts was very low. Regarding PFHxA, the -COOH group was targeted by electrons, giving rise to the creation of unstable perfluoroalkyl radicals through an electron-mediated chain initiation process.

**Table 1**
Summary of the byproducts with proposed structures, retention time (RT), molecular formula, *m/z* value.

| Parent compound (matrix type) | Byproduct TP mark | Name | Molecular formula | RT | Observed mass *m/z* | Structure |
|---|---|---|---|---|---|---|
| PFOS (TW) | TP_448 | 1,1,2,2,3,3,4,4,5,5,6,6,7,7,7-pentadecafluoroheptane-1-sulfonic acid (PFHpS) | $C_7HF_{15}O_3S$ | 9.87 | 449.9333 | 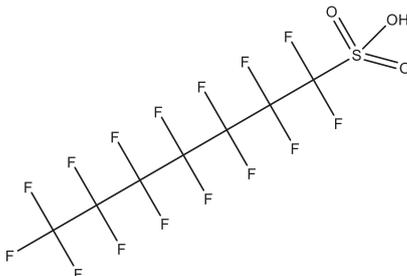 |
| PFOS (TW) | TP_212 | 2,2,3,3,4,4,4-heptafluorobutanoic acid (PFBA) | $C_4HF_7O_2$ | 4.89 | 212.9793 | 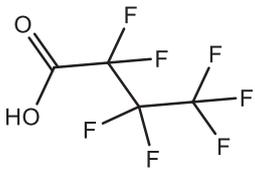 |
| PFHxA (DW) | TP_246 | 2,2,3,3,4,4,5,5,5-nonafluoropentanal | $C_5HF_9O$ | 4.80 | 246.9815 | 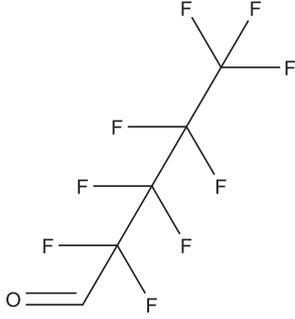 |
| PFBA (TW) ADONA (TW) | TP_157 | 2,2,3-trifluoro-3-methoxypropanoic acid | $C_4H_5F_3O_3$ | 0.54 | 157.0117 | 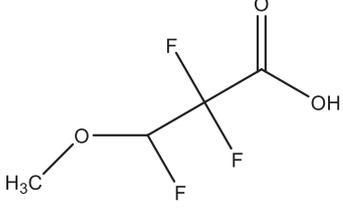 |
| ADONA (DW, TW) | TP_250 | 1,1,2,2,3,3-hexafluoro-3-(trifluoromethoxy)propan-1-ol | $C_4HF_9O_2$ | 6.35 | 250.9758 | 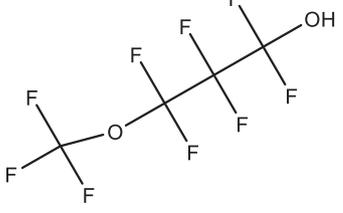 |
| ADONA (TW) | TP_105 | 2,3-dihydroxypropanoic acid | $C_3H_6O_4$ | 0.86 | 105.0192 | 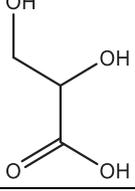 |

After 10 min of reaction in TW, a byproduct identified as TP_157 (*m/z* 157.0117) originating from PFBA was detected. This byproduct arises from H/F exchange process. Moreover, the C-C breaks within the molecule can interact with •OH through intramolecular reaction, ultimately resulting in the formation of an ether compound. The same TP was identified in ADONA samples in TW, which correspond to C-C breaks and defluorination. Since both TW and SE are complex matrices, electrons and ions needed for degradation of PFAS have less ability to react with target compound due to the presence of scavengers such as $NO_3^-$. This could be a possible explanation for why the same byproducts are not found in different solutions at the same treatment time. Rest of the byproducts that were found in other solutions were TP_246 DW in case of PFHxA and was defined as radical form which was present in 3–10 min treatment time, with slowly decreasing trend after 3 min. Attack of the electrons or argon ions and the OH radical resulted with PFHxA molecule losing (-$CF_2O$) group.

*3.3.2. Ether byproducts*

The identification of byproducts, namely TP_250 (*m/z* 250.9758) and TP_105 (*m/z* 105.0192), originating from ADONA, has been established. TP_250 was detected in DW within the 3–7-min treatment, though its presence was no longer detectable after a 7-min treatment. Conversely, in TW, TP_250 persisted from 3 min onward, with a slight decrease noted until the conclusion of the treatment. Additionally, the presence of TP_105 was noted in TW, with its concentration progressively increasing between 3 and 10 min of treatment time.

These results point to several possibilities: *i)* same types of byproducts can be found in different cases, but the higher the complexity, it takes longer times for the detection/formation of byproducts; *ii)* more than one reaction is possible during treatment time; *iii)* level of byproduct concentration is too low in the certain treatment time and does not exclude the presence of the byproduct; *iv)* in DW, byproduct was also degraded during treatment since there was no complex interferences with reactive species. Fig. 5a and b compare ion intensity trend of ADONA as parent compound with detected byproducts in DW and TW.

It was found that a possible byproduct in GenX sample was trifluoroacetic acid TFA ($C_2F_3O_2H$). But since the same byproduct has been detected in some blank samples, it was impractical to conclude whether TFA is a product or was trapped in the system in MS. Therefore, in this paper the proposal of degradation has been given only for ADONA (Fig. 6) with the verity that the same is possible for GenX based on similar structure of those compounds.

While the carboxylic or sulfonic group is favoring site for direct electron transfer, causing a cleavage of C-C or C-S bond in other PFAS, here, oxidative degradation occurs which causes split of ether-bond (C-O cleavage). With further reactions of decarboxylation, defluorination, HF exchange, TP_250 is converted into TFA. Based on the structure of parent compound and byproducts, it can be seen that C-O cleavage is on the different carbon for TP_157. As for TP_105, it has been defined as a third possible byproduct.

Based on the elucidation of byproducts resulting from PFAS treatment using CAP, can be outlined that the degradation of PFAS in water occurs through electron transfer, giving rise to the formation of corresponding radical anions which means that reactive species trigger chain reactions leading to the cleavage of C-F, C-C, or C-S bonds within PFAS molecules.

**4. Conclusion**

The results obtained in this work demonstrate the success of application of non-thermal plasma at atmospheric pressure (NTP-APPJ) for PFAS- contaminated water treatment on laboratory scale, for three different water matrices. In DW, high removal efficiency (>90%) was achieved for longer chain PFAS; PFOS, PFHxA and ADONA in 10 min of treatment time while other compounds showed less removal efficiency, but still high for short treatment time. In TW and SE which represent real water samples, overall, percentage of degradation was lower, between 8% and 50%, depending on the compound and matrices. However, while the short chain PFAS in SE showed almost the same percentage as in TW, PFOS was around 16% and GenX and ADONA 39% and 49%, respectively. It can be concluded that degradation is compound-depending and influenced by different matrix effects. What is more important is that the novel compounds, PFAS substances ADONA and GenX have shown promising results and successful removal in plasma system that was used in this work. Overall, results showed that plasma can be successfully used for the treatment of PFAS contaminated water, especially for the novel compounds, ADONA and GenX which are in production and used up to this date.

The focus of this study was also investigation of byproducts produced utilizing plasma-based technology, especially ADONA and GenX. In total, 7 different byproducts were detected. The results proved one of previous hypothesis, that, in case of PFSAs and PFCAs compounds, electron transfer was important initiation step of PFAS degradation, breaking of the C-C bond or C-S bond on the head group. Degradation was then followed by decarboxylation, defluorination, hydrolysis and reaction with OH radical. As for ADONA and GenX, initial break of the bond is on the ether group, possibly followed by process of

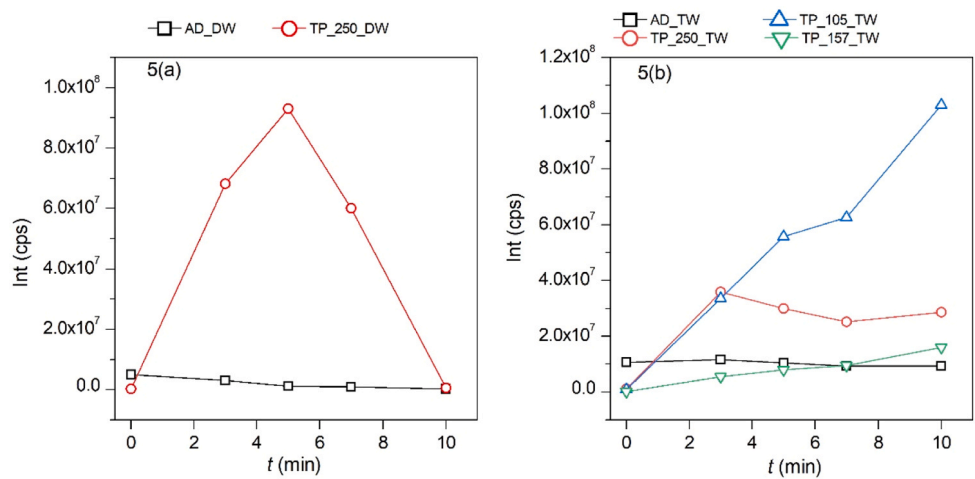

**Fig. 5. (a) and (b).** Ion intensity (counts per second - cps) of MS signals of ADONA and byproducts: TP_250 (red line/circle) in DW (5a) and TP_250 (red line/circle), TP_105 (blue line/upper triangle) and TP_157 (green line/down triangle) in TW (5b).

...

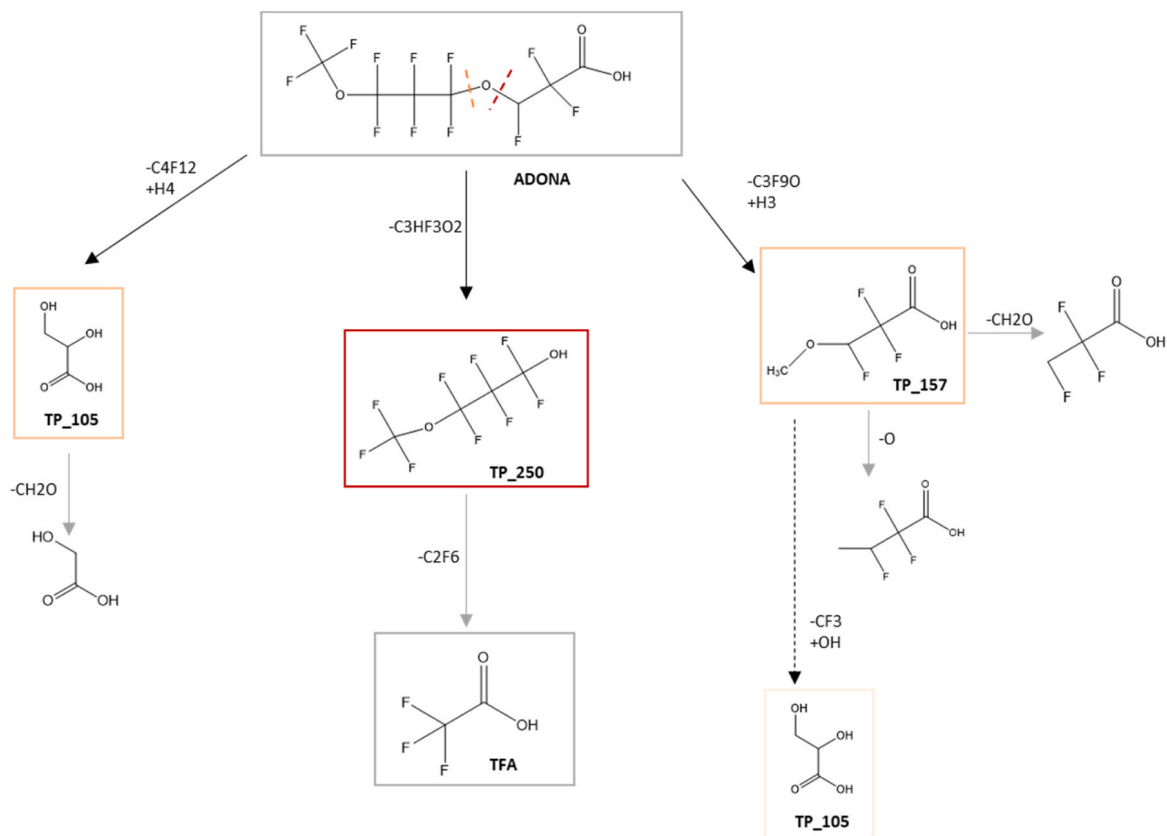

**Fig. 6.** Identified products of ADONA and possible degradation pathway.

defluorination and decarboxylation. It can be concluded that both reduction and oxidation processes are involved in PFAS degradation, and the mechanism can vary in different compound types. Therefore, application of plasma seems to be a good approach, but more investigation is needed for attaining better performance with the aim to achieve complete mineralization which is needed for PFAS case.

A degradation mechanism of ADONA is proposed in this paper. According to authors knowledge, this was the first attempt to investigate and propose the degradation of ether-group PFAS, ADONA and GenX in plasma-based processes. Therefore, further investigation is needed to demonstrate detailed pathway. Authors also suggest further investigation in terms of toxicity of newly investigated intermediates.

**CRediT authorship contribution statement**

**Elisabeth Lumbaque:** Writing – review & editing, Formal analysis. **Nikola Skoro:** Writing – review & editing, Validation, Resources. **Nevena Puac:** Writing – review & editing, Validation, Resources. **Olivera Jovanovic:** Investigation, Formal analysis. **Barbara Topolovec:** Writing – original draft, Visualization, Validation, Resources, Methodology, Investigation, Formal analysis, Conceptualization. **Mira Petrovic:** Writing – review & editing, Validation, Supervision, Resources, Project administration, Funding acquisition.

**Declaration of Competing Interest**

The authors declare that they have no known competing financial interests or personal relationships that could have appeared to influence the work reported in this paper.

**Data Availability**

Data will be made available on request.

**Acknowledgements**

This work was supported by the European Union's Horizon 2020 research and innovation programme under the Marie Sklodowska-Curie grant agreement – MSCA-ITN-2018 (grant number 812880). ICRA researchers thank funding from CERCA program.

**Appendix A. Supporting information**

Supplementary data associated with this article can be found in the online version at doi:10.1016/j.jece.2024.112979.